# Linking Software Development and Business Strategy Through Measurement

Victor Basili[B,C], Jens Heidrich[A], Mikael Lindvall[B], Jürgen Münch[A],
Myrna Regardie[B], Dieter Rombach[A,E], Carolyn Seaman[B,D],
Adam Trendowicz[A]

[A] Fraunhofer Institute for Experimental Software Engineering (IESE),
Kaiserslautern, Germany
Tel. +49-631-6800-0
{jens.heidrich | juergen.muench | dieter.rombach | adam.trendowicz}@iese.fraunhofer.de

[B] Fraunhofer Center for Experimental Software Engineering (CESE),
College Park, MD, USA
Tel. +01-301-403-0
{basili | mikli | mregardie | cseaman}@fc-md.umd.edu

[C] University of Maryland, College Park, MD, USA
[D] University of Maryland Baltimore County, Baltimore, MD, USA
[E] University of Kaiserslautern, Kaiserslautern, Germany

## Abstract

Most of today's products and services are software-based. Organizations that develop software want to maintain and improve their competitiveness by controlling software-related risks. To do this, they need to align their business goals with software development strategies and translate them into quantitative project management. There is also an increasing need to justify cost and resources for software and system development and other IT services by demonstrating their impact on an organization's higher-level goals. For both, linking business goals and software-related efforts in an organization is necessary. However, this is a challenging task, and there is a lack of methods addressing this gap. The GQM⁺Strategies® approach effectively links goals and strategies on all levels of an organization by means of goal-oriented measurement. The approach is based on rationales for deciding about options when operationalizing goals and for evaluating the success of strategies with respect to goals.

## Need for Business Alignment

Along with the growth in society's dependence on software and other forms of information technology (IT), the size and complexity of software systems have also grown. This has only magni-fied the cost, schedule, and quality concerns that have always plagued software development efforts. For decades, software engineering researchers and practitioners have attempted to control and reduce the costs of building software, to produce working software within shorter periods of time, and to increase the quality of the software produced. While great strides have been made in all three areas, the growth of software, along all dimensions (size, complexity, pervasiveness, criticality, etc.), has outpaced our ability to control all the factors related to its development.

What has become clear, however, is that the issues related to software cost, schedule, and quality are inextricably linked with larger issues facing the businesses that develop the software.

Such businesses come in a variety of flavors. Some are in the business of selling the software they develop to customers, either as custom-built software on contract or shrink-wrapped applications for some segment of the population. Others are in the business of selling some product or service, of which software is a significant component. Still others may only develop software to support their internal IT infrastructure, and do not sell software-related products. Some software organizations are not in business at all, but are non-profit organizations, government entities, or educational institutions. While all these organizational configurations provide quite different challenges to their development projects, the key here is that all software is developed within a larger business context, encompassing larger business goals, strategies, and measures of success.

While all businesses employ various strategies to achieve their objectives, these objectives are not always stated explicitly or clearly enough to allow one to check whether or not they are achieved. Further, how these objectives are translated into lower levels of the business and into individual projects is often even less clear. A methodology is needed to bridge the gap between business strategies and their implementation at the project level.

To understand the relationships between the business and project level goals, and to verify their achievement, quantitative data is required. This is one reason why quantitative measures are required at high-maturity software organizations. But software improvement strategies, such as CMMI and ITIL, are not directly linked to business value. Such a linkage must be made explicit or the investment in collecting data does not result in the expected benefits, and the contribution of project performance to the achievement of strategic goals remains unclear. This has the practical effect of the software measurement effort losing its support among decision makers in the organization, and thus eventually failing.

The popular Goal Question Metric (GQM) approach[1] has served the software industry well for several decades in defining measurement programs. However, it does not provide *explicit* support for motivating and integrating measurement at various levels of the organization, e.g., project goals, business goals, strategies, and assumptions. On the other hand, approaches such as Balanced Scorecard[2] address mainly business-level goal-setting activities, and do not support the alignment of objectives at different levels of the organization with an integrated methodology. Approaches such as CoBIT[3] are very focused on a particular application field (such as IT gover-nance) and define a very detailed model that fits the application field quite well, but has to be followed strictly in order for the approach to be applied. The gap in current practice is the lack of explicit linkages between different levels of the organization and the flexibility to adapt and tailor the approach to the specific needs and objectives of the organization.

To fill this gap, we propose GQM⁺Strategies® (registered trademark application pending): an integrated approach that is based on the popular GQM approach and adds the capability to create measurement programs that ensure alignment between goals and strategies at different levels, from the highest strategic levels of the business to the level of individual development projects. The approach is derived from experiences in the software domain, but it is also intended to be applicable in the systems domain.

In what follows, we will look at the foundations of the approach by addressing related work and its influence on GQM⁺Strategies®. Next, we will illustrate the application of GQM⁺Strategies® to an example business goal, summarize its benefits and features, and describe how the resulting model would be used to efficiently implement a business strategy at all levels of the organization.

## Background

As background for this work, we discuss various approaches to goal-oriented software measurement, one of which (GQM) is the basis of the work described in the rest of the paper. Common problems that software development organizations encounter in instituting measurement programs include too much data collected, not the right data collected, and insufficient analysis of the data that is collected. This leads to numerous problems, including decreased cost effectiveness of the measurement program and disillusionment about metrics on the part of developers and managers. The end result is often the eventual failure of the measurement program as a whole.

In response to such problems, several structured approaches to software measurement have been developed and are used in organizations. These approaches are referred to as "goal-oriented" approaches because they use goals, objectives, strategies, or other mechanisms to guide the choice of data to be collected and analyzed in a systematic way.

The GQM approach[1] provides a method for an organization or a project to define goals, refine those goals into specifications of the data to be collected, and then analyze and interpret the resulting data with respect to the original goals. Implicit in the GQM approach is the use of interpretation models. These models help practitioners interpret the data yielded by the metrics in the specific context.

Balanced Scorecard[2] (BSC) links strategic objectives and measures. The "scorecard" consists of four perspectives: financial, customer, internal business processes, and learning and growth. BSC can be viewed as a tool for defining strategic goals from multiple perspectives beyond the purely financial focus. BSC does not dictate a static set of measures, but serves as a framework for strategic measurement and management.

Practical Software Measurement[4] (PSM) offers detailed guidance on software measurement, including a catalogue of specific measures and information on applying them in an organization. PSM also includes a process for choosing appropriate measures based on the issues and objectives relevant to a software development project.

To address the issue of aligning business-level and software-level measurement, various combinations of BSC, GQM, and PSM have been proposed[5, 6, 7, 8]. Although these approaches recognize the need to link organizational goals to lower-level goals, they do not recognize or support truly *different* goals at different levels of the organization that are linked explicitly, enabling us to feed the analysis results and interpretations back up the chain. In our approach, we create mappings between the data related to goals at different levels, so that insights gained relative to one goal at one level can still support and contribute to satisfying goals at higher levels, without requiring that each level share the same goals.

For specific domains and application fields, specific approaches exist that emphasize the need for linking business goals to lower-level properties. For instance, there is an increasing awareness that the IT infrastructure itself imposes significant risks on a company. As a consequence, several regulatory constraints in the IT governance domain and the IT service domain, such as Sarbanes-Oxley Act[9] (SOX), have been developed recently. The solutions proposed by models in these domains, especially CoBIT® 4.1[3] and ITIL release 3, offer connections between predefined sets of goals and attributes of the IT infrastructure. CoBIT, for instance, is based on a process model that subdivides IT into four domains and 34 processes. However, there is no mechanism for adapting and tailoring the solution to the specific needs of an organization, addressing the specific context as well as documenting inherent assumptions. There is no clearly defined interpretation model that indicates if an overall strategy is working or has to be changed to avoid business failure.

## GQM⁺Strategies®

GQM⁺Strategies® is a measurement planning and analysis approach, the output of which is a detailed and comprehensive model that defines all the elements necessary for a measurement program. In extending GQM, the GQM⁺Strategies® approach makes the business goals, strategies, and corresponding lower-level goals explicit. Strategies are formulated to deal with business goals such as improving customer satisfaction, garnering market share, reducing production costs, and more, taking into account the context and making explicit any assumptions. Strategies also help define lower-level goals that can be assigned to different parts of the organization, e.g., software goals, hardware goals, marketing goals, etc. Again, strategies may be formulated to deal with lower-level goals. The number of goal/strategies levels depends on the (internal) structure of an organization. For our work, we mainly focus on software-related goals at the second level because we are concerned with relating software project measurement to higher-level business goals. GQM+Strategies® also makes explicit the relationships between goals/strategies and measurement goals. Measurement goals are broken down into concrete metrics using the GQM approach. Interpretation models (based on the metrics) are defined for determining whether a strategy was successful and a related goal could be achieved. Attached to goals and strategies at each level of the model is information about relevant context factors and about assumptions (defining the rationale for choosing specific goals and strategies). The entire model provides an organization with a mechanism for not only defining measurement consistent with larger, upper-level organizational concerns, but also for interpreting and rolling up the resulting measurement data at each level. GQM+Strategies® linkages and measures ensure that the business goals are fulfilled.

Rather than provide a step-by-step tutorial on the method for defining a GQM+Strategies® model[10], in this article we will instead present a completed model from an example organization, and then use the example to illustrate the features and benefits of the GQM+Strategies® approach.

## *Example Application*

Our example application of GQM+Strategies® takes place in an organization, which we will call ABC, that provides information services to customers through the Web. That is, customers pay a service fee for access to information and to software that searches, analyzes, and presents that information. Customers do not pay for the software itself. Thus, the business model implies that the amount of revenue generated is determined by the number of times customers access the ABC system. A representation of the GQM+Strategies® model for this example is presented in Figure 1. The whole model is described step by step below.

The starting point of the GQM+Strategies® process is a **business goal**. In this example, one of ABC's business goals is to increase profit from software service usage (Goal 1 in Figure 1). The GQM+Strategies® approach enforces the explicit documentation of the relevant **context factors** and **assumptions** that are necessary for understanding and evaluating each goal. In the case of this business goal, one such context factor is that the amount of revenue generated at ABC is determined by the number of times customers access the ABC software products. Other details are documented in the GQM+Strategies® business goal template, shown in Figure 2, and include the desired **magnitude** of the improvement, the **timeframe** for achieving the goal, the **scope** including the organization and the individual primarily responsible for achieving the goal, and any **constraints** or conflicting goals. **Relations** to other goals are also documented in the template.

There is an assumption here that there are enough projects with a CMMI maturity level greater than 1 such that if just those projects provide a 15% improvement, the organization can manage a 10% improvement overall. Associated with each goal in the model is a measurement and evaluation framework, based on GQM, specifying how the goal should be evaluated, what data needs to be collected, and how that data should be interpreted. The nodes of each **GQM graph** (i.e., each green box on the right side of Figure 1) consist of a **measurement goal**, which describes what knowledge needs to be gained from the measurement activity, a set of **questions** to be answered, the **metrics** and data items that are required to answer the questions, and an **interpretation model** that specifies how the data items are to be combined and provides the criteria for determining the success of the goal. These nodes are related in a semi-hierarchical fashion. Each goal on the left side of the model may have several associated measurement goals, each of which is the basis for an entire GQM graph. However, it is expected that different GQM structures will use some of the same questions and metrics, and interpretation models may combine data from different GQM structures, thus optimizing the metrics collection process.

In the example, the GQM graph that defines how the business goal is to be evaluated is based on the **measurement goal** (the gold box labeled G1), which, in full GQM notation[1], would be: <u>Analyze</u> the trend in profit <u>for the purpose of</u> evaluation <u>with respect to</u> a 10% increase in annual income per year <u>from the point of view</u> of ABC's management <u>in the context of</u> the ABC organization. This goal leads to the following questions (Q1 and Q2 in Figure 1): What is the current profit (measured by $P_0$)? What is the profit for each succeeding year (as measured by $P_x$)? The results can be analyzed using decision criteria incorporated into the **interpretation model** (top white box in Figure 1). The model says that, starting in year 2, if the profit for the current year ($P_2$) is at least 10% (i.e., 1.1 times) higher than the profit for the initial preceding year ($P_1$), then the goal has been satisfied. The full interpretation model (not shown in Figure 1) also includes an "else" part related to the effectiveness of the chosen strategies, which is explained below**.**

Associated with each goal is a **strategy** (in the lower part of the upper left-hand green box in Figure 1, labeled Strategy 1), which the GQM+Strategies® user must enumerate and then choose from among a set of potential strategies, taking into account various influencing context factors. In the case of the ABC business goal, possible strategies for meeting the business goal might be to deliver added capabilities to encourage more usage, increasing the rates charged to customers, or reducing development costs. It was decided to follow a strategy of delivering added functionality in the product releases at regular and frequent intervals. An **assumption** that must be made explicit at this point is that added functionality will lead to increased customer satisfaction, which will in turn lead to higher usage. The combination of a goal and a strategy (i.e., each of the green boxes on the left side of Figure 1) is called a GQM+Strategies® element.

At the next lower level of the model, we address a goal that is derived from the strategy (or strategies) chosen at the top level. The second goal (Goal 2, in the second green box on the left side of Figure 1) is to deliver a new release of the software each six months that incorporates at least 5% more functionality than the previous release, with the new functionality coming from the backlog of customer-requested requirements, and to keep the cost of each release within 10% of the estimated cost. At this level, the goal is specific to the software development portion of the organization, so we refer to this goal as a **software goal**. In general, the name of the second level depends on the organization in which GQM+Strategies® is applied and the overall number of levels that have to be modeled. Part of this software goal is depicted in Figure 3 in the GQM+Strategies®' goal template.

Another software goal, related to the 10% variance in cost, would be described by a similar template. The software goal template asks for the same categories of information as the business goal template, but allows for the measurement and interpretation model to be further refined and linked to the higher-level goals.

At this point, the interpretation model for the business goal (topmost white box in Figure 1) can be further refined by adding information to the "else" part, as shown in Figure 4. Note that the full interpretation is dependent on the lower-level goals, e.g., if the functionality was not increased by 5%, then the level 2 strategy was not effective, etc. The interpretation model becomes more and more detailed, with more conditional logic, at each lower level in the GQM+Strategies® model.

It is the responsibility of the software development organization to develop and carry out a strategy for accomplishing this software goal. The strategy chosen in this example (Strategy 2 in Figure 1) is to adopt an approach to rating the importance of different requirements, like the MoSCoW approach[12] for requirements and release planning, and to adopt COCOMO[11] for cost estimation. A *context factor* relevant to this step in the process (and which must be documented explicitly) is that there is an expert available who knows and recommends the MoSCoW approach, but there is no experience with this method at ABC.

Three *assumptions* that are relevant here are that the organization (1) can estimate the percentage of functionality delivered, e.g., it can use a proxy such as additional lines of code delivered, number of function points delivered, or a formula based upon a count of actual requirements; (2) that the difficulty and importance of requirements are weighted in some way (hard, medium, easy) to provide input to the cost model; and (3) that the backlog of customer-requested requirements continues to grow.

As with the business goal at the top level of the model, there is a set of GQM structures that define how the software goal will be evaluated (the second green box on the right side of Figure 1). The GQM goal (G2 in Figure 1): <u>Analyze</u> each 6-month release <u>for the purpose of</u> evaluation <u>with respect to</u> incorporation of 5% new functionality as compared to the previous release <u>from the point of</u> view of the Web services project manager <u>in the context of</u> the ABC organization. The questions (Q3 and Q4) include: What was the amount of functionality delivered at each release? What was the percentage of new requirements of different importance included? Was each release delivered within six months of the previous one? The interpretation of the achievement of this software goal is: **If**, at each 6-month milestone, the growth in functionality of a release ³ 5%, **then** the level-2 goal is satisfied for this release, **else** assumptions about MoSCoW are incorrect **or** we have not chosen the correct strategy. We can also further refine the interpretation model at the business goal level, e.g., **if** the business goal is satisfied (meeting 10% increase in profit), but goal 2 is not, **then** assumptions are wrong (e.g., delivery of some particular requirement alone caused the gain).

The last level of goals shown in Figure 1 starts with a goal that is derived from the strategy above, and which is applicable to a particular software development group or project. The goal at this level (Goal 3 in Figure 1) is to apply the MoSCoW and COCOMO approaches effectively. The relevant development group has developed a strategy for this goal that involves training personnel, acquiring tools that will assist in the application of these methods, and piloting the methods on a single project. A relevant *assumption* that should be documented at this point is that training for these approaches can be targeted to a few specific individuals, so the impact of the training on cost and schedule is reasonable. The GQM graph at this level, only part of which is shown as the bottom right-hand box in Figure 1, involves evaluating the effectiveness of the use of MoSCoW and COCOMO as well as the training and tools used. The results of this interpretation are then included in the interpretation of the higher-level goals. Some of the questions and metrics defined at the second level can be reused at the third level. This efficiency due to reuse is a benefit of the GQM approach, which is inherited by the GQM$^+$Strategies® approach.

Using this example, we have shown the results of applying GQM$^+$Strategies® in a single context. This example shows three levels of goals, but there might be many more levels in other situations. Generally, however, the process begins at the business goal level, passes to lower-level goals, and then, finally, to project-specific goals. Each of these layers may have several levels of goals, and there may be multiple peer goals at each level. But each level must be based on well-defined goals, each with an associated strategy, documented context factors and assumptions, and a measurement and evaluation framework (including measurement goals, questions, metrics, and interpretation models) defining the evaluation of the goal. In addition, conflicts and relationships between goals can be documented and tracked with GQM$^+$Strategies®, although this was not shown in this example. The next sections describe what an organization can do with such a model and the benefits of developing and using it.

## *Support for Establishing Strategic Measurement*

The GQM$^+$Strategies® approach provides a number of features for organizations who want to create a software measurement program that is consistent with and contributes to the achievement of goals at all levels of the organization. This is achieved through explicit linkages between goals at the strategic level, the software development level, and the operational project level. This linkage is achieved through the specification of strategies, as we see in our example in Figure 1. The first (increasing profit) and second (new functionality in short releases) goals are linked through a strategy that specifies that increased profit will be achieved by providing cus-

8 of 16

tomers with more functionality. Often this linkage is implicit in organizations, but making it explicit has many benefits.

Templates are provided to help define all types of goals at the level of detail necessary and to track their relationships to each other. In the example, the full template for the business goal would include information such as the target increase in profit, the timeframe, the responsible parties, and any constraints or conflicting goals. Templates have been developed for all types of goals in the model.

The approach also has a built-in capability for tracking context factors and assumptions at each level, so that the effect of changes in the context and the status of the assumptions can be assessed more easily. In the example, the approach requires that the assumption about the training required for MoSCoW and COCOMO be documented, so if at some later time the assumption turns out to be false, the model will indicate what elements are affected by that assumption, and most likely will have to be re-evaluated.

Another feature of GQM+Strategies® is the use of interpretation models, which tie together measurement goals, context factors, assumptions, and data in a model that facilitates correct and useful interpretation of the results of measurement. This idea is borrowed from the original GQM approach, but is broadened here to allow interpretation models at each level to inform not only that level, but also higher levels as the data is aggregated and rolled up. In our example in Figure 1, the results of applying the interpretation model at the lowest level will yield information about how the piloting of MoSCoW and COCOMO went, as well as information about the training and tools. The information from the lower level can help in the diagnosis of any problems encountered at the next level up. Further, the results of the interpretation at the second level can inform the analysis at the top level. At the top level, if profit does not increase as expected, then the analysis results from the second level will help determine whether the problem is due to higher costs, inadequate functionality delivered, late releases, or some other cause.

The entire approach is further supported by an experience base, which can be instantiated with an organization's own past measurement experience, or which can be used as is, populated with the considerable experience of the creators of the approach.

The GQM+Strategies® method distinguishes between eight conceptual elements that form the basis for constructing a consistent model (see sidebar). Figure 5 gives an overview of all conceptual components for constructing a consistent model. These components allow multiple goal levels and permit multiple strategies for each of these goals. Strategies also help define lower-level goals that can be assigned to different parts of the organization, e.g., software goals, hardware goals, marketing goals, etc. A set of predefined goals and strategies may become part of an (organization-specific) experience base. Context information about the organization and assumptions are drivers for this instantiation process and influence the definition of new goals and strategies as well as the selection and adaptation process for predefined goals and strategies. At each level of the instantiation process, GQM plans are defined in order to measure the achievement of the defined goal in combination with the chosen strategy. This includes the definition of GQM measurement goals, the derivation of questions and metrics, as well as the definition of an interpretation model that determines whether the measurement goal has been reached.

## *Conclusion: Benefits of Applying GQM+Strategies®*

The most important benefit of applying GQM+Strategies® is the resulting transparency of measurement motivations and goals at different levels of the organization. This allows the identification of goal relationships and conflicts. It also facilitates improved, more effective communication between the business and software segments of an organization. In the ABC example, the GQM+Strategies® model helps project personnel planning their implementation of MoSCoW and COCOMO to understand why they have been asked to implement these approaches. This might result in more focused training or tailoring of COCOMO, which makes it more applicable to very short development cycles. Without the GQM+Strategies® model, project personnel might misdirect their training efforts, or even choose the wrong tools.

Other existing software measurement approaches and business strategy approaches can be integrated into GQM+Strategies®, as it is a very flexible method. For example, Balanced Score Card[2] might be used at ABC to define business goals and even strategies, which can then be used as a starting point for GQM+Strategies®. On the other hand, PSM[4] would be useful in defining the low-level definitions of the metrics listed in the right-hand column of Figure 1.

Other benefits of GQM+Strategies® include those of software measurement in general, but raised to the organizational level, rather than the individual project level or even the software part of the organization. Sharing measurement planning and results across the organization results in lower costs (and better ROI) of measurement, increased success of software measurement programs, better risk identification and management, and compliance with SPI models (e.g., CMMI). This comes in large part from linking the interpretation models at each level with those at levels above.

Finally, one substantial corporate benefit from using GQM+Strategies® is the ability to build a corporate measurement experience base. Such an experience base can become a valuable corporate asset that facilitates project measurement and planning over time and lowers project costs. Such an experience base can begin with the set of generic experiences already in our base, but becomes more valuable as it is instantiated with more and more organization-specific models. For example, any part of the model presented in Figure 1 can be reused later. The assumptions and context factors are particularly important in this case, as they capture the properties that might not hold in another situation in which the model might be reused. But because the assumptions and context factors are captured explicitly and are linked to particular goals, strategies, etc., it is clear then which parts of the model need to be re-evaluated when an assumption or context factor changes. Finally, this allows organizations the flexibility to adapt their goals and strategies to market needs and analyze the consequences for their organization. The approach is currently being used to define goals and strategies for different types of organizations.

# Biographies

**Victor R. Basili** is a professor at the University of Maryland and chief scientist of the Fraunhofer Center for Experimental Software Engineering (CESE), USA. His research interests include measuring, evaluating, and improving the software process and product via empirical studies.

**Jens Heidrich** is a researcher at Fraunhofer Institute for Experimental Software Engineering (IESE), Germany. His research interests include project management, quality assurance, and measurement. He received his MS in computer science from the University of Kaiserslautern, Germany.

**Mikael Lindvall** is a senior scientist and division director at Fraunhofer Center for Experimental Software Engineering (CESE), USA. His interests include agile methods, software process improvement, software architectures, and experience and knowledge management. He received his PhD in computer science from Linköping University, Sweden.

**Jürgen Münch** is division manager for quality management at the Fraunhofer Institute for Experimental Software Engineering (IESE), Germany. His research interests include quality assurance, process management, and measurement.

**Myrna Regardie** is a senior engineer at Fraunhofer Center for Experimental Software Engineering (CESE), USA. Her research interests include software process improvement, measurement, project management, and knowledge management. She received her BS in mathematics from Juniata College, Huntingdon, USA.

**Dieter Rombach** is a professor at the University of Kaiserslautern, Germany. He holds a chair in software engineering, is executive director of the Fraunhofer Institute for Experimental Software Engineering (IESE), Germany, and chairs the Fraunhofer Information and Telecommunication Technology (ICT) group.

**Carolyn Seaman** is an associate professor of information systems at the University of Maryland Baltimore County and a scientist at Fraunhofer Center for Experimental Software Engineering (CESE), USA. She received her PhD in computer science from the University of Maryland, USA.

**Adam Trendowicz** is a researcher at Fraunhofer Institute for Experimental Software Engineering (IESE), Germany. His research interests include software cost modeling, measurement, and process improvement. He received his PhD in computer science from the University of Kaiserslautern, Germany.

# Sidebars

**Sidebar: GQM⁺Strategies® Terminology**

| | |
|---|---|
| **Business Goals:** | Goals the organization wishes to accomplish in general in order to achieve its strategic objectives. |
| **Context Factors:** | Environmental variables that represent the organizational environment and affect the kind of models and data that can be used. |
| **Assumptions:** | Estimated unknowns that can affect the interpretation of the data. |
| **Strategies:** | A set of possible approaches for achieving a goal that may be refined by a set of concrete activities. |
| **Level i Goals:** | A set of lower-level goals inherited from level i-1 goals as part of the level i-1 goal strategy, e.g., a goal related to a project that is part of the software strategy decision. |
| **GQM Goals:** | Goals defined so that they can be measured using the GQM approach. A GQM goal is associated with goals at all levels. |
| **Interpretation Models:** | Models that help interpret data to determine whether goals at all levels are achieved. |
| **GQM⁺Strategies® Element:** | A single goal and derived strategies (including a set of concrete activities), as well as all context information and assumptions that explain the linkage between the goal and corresponding strategies. |
| **GQM Graph:** | A single GQM goal (that measures a GQM⁺Strategies® Element), corresponding questions, metrics and interpretation models. |

# Figures

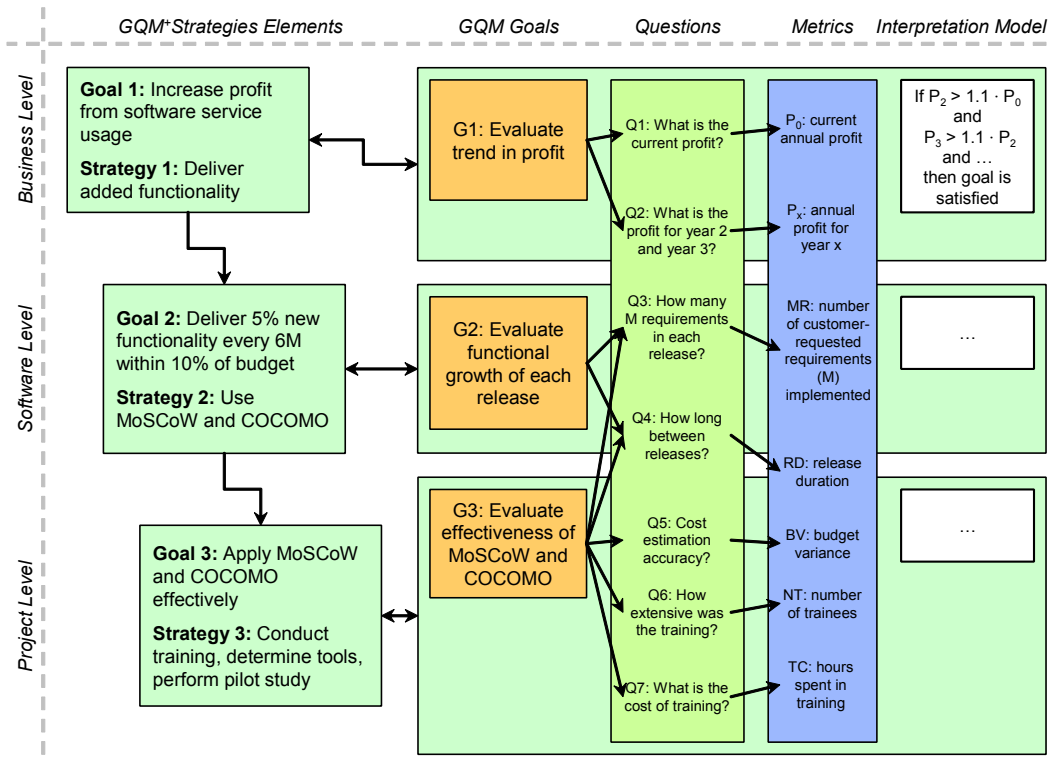

**Figure 1: A GQM+Strategies® Model for ABC**

| Activity:    | Increase                                              |
|--------------|-------------------------------------------------------|
| Focus:       | Net Income                                            |
| Object:      | ABC Web Services                                      |
| Magnitude:   | 10% per year                                          |
| Timeframe:   | Annually, beginning in 2 years                        |
| Scope:       | Development groups assessed at CMMI level 2 or higher |
| Constraints: | Available resources, ability to sustain CMMI levels   |
| Relations:   | CMMI-related goals                                    |

**Figure 2: Business Goal for ABC**

| | |
|---|---|
| **Activity:** | Deliver |
| **Focus:** | More usable functionality, e.g., M (must) type requirements from the backlog of customer-requested requirements |
| **Object:** | Each release of ABC Web Services software |
| **Magnitude:** | 5 % more functionality than the prior release |
| **Timeframe:** | Every 6 months, beginning in 2 years |
| **Scope:** | Web services development projects with CMMI level 2 or higher |
| **Constraints:** | Available resources, ability to sustain CMMI levels, ability to estimate cost and schedule for a release |
| **Relations:** | Achievement of cost and schedule estimate accuracy, ability to improve CMMI levels of development groups |

**Figure 3: Software Goal for ABC**

**for** x = 2, 3, …,
**if** $P_x \geq 1.1 * P_{x-1}$ **then**
    the goal has been satisfied,
**else if** functionality was increased appropriately, **then**
    either some assumption is incorrect or we have chosen the wrong level 1 strategy.

**Figure 4: Extended Interpretation Model**

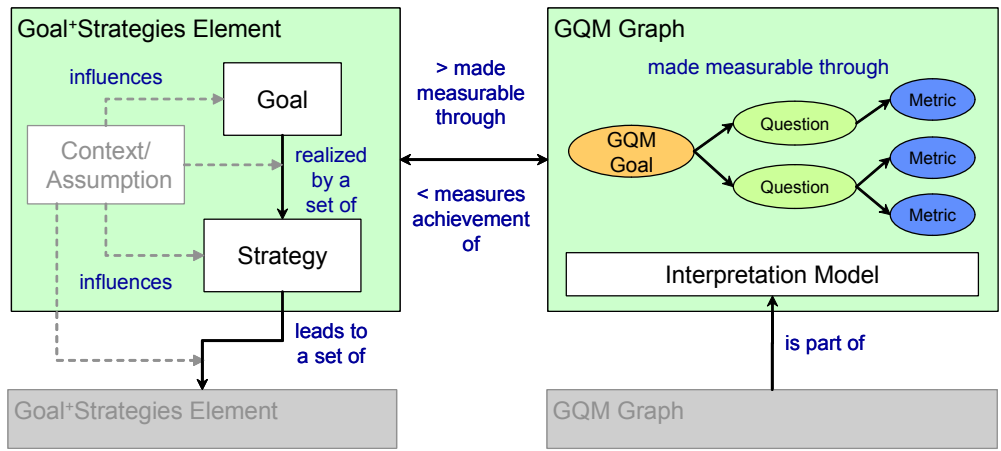

**Figure 5: GQM+Strategies® Components**